%% file: gamma-ichep2010.tex
\documentclass{PoS}

\title{Measurements of the CKM angle $\gamma$ at \babar}

\ShortTitle{Measurements of the CKM angle $\gamma$ at BABAR}

\input babarsym
\input mysymbols

\author{\speaker{Fernando Mart\'{\i}nez-Vidal}\thanks{On behalf of the \babar\ Collaboration.}\\
   Instituto de F\'{\i}sica Corpuscular (IFIC), Universitat de Val\`encia-CSIC,\\
   Apartado de Correos 22085, E-46071 Valencia, Spain \\
        E-mail: \email{fernando.martinez@ific.uv.es}}

\abstract{
The recent measurements of the CKM angle $\gamma$ by the \babar\ experiment are reported. 
The analyses have been performed using the complete sample of 468 million \BB pairs collected by
the \babar\ detector at the SLAC \pep2 asymmetric-energy \B factory during the years 1999-2007.
}

\FullConference{35th International Conference of High Energy Physics - ICHEP2010,\\
		July 22-28, 2010\\
		Paris France}

\begin{document}



In the standard model (SM) of particle physics, \CP violation in the quark sector of weak interactions arises from a single
irreducible phase in the Cabibbo-Kobayashi-Maskawa (CKM) that describes the mixing of quarks~\cite{ref:CKM}. 
The unitarity of the CKM \mbox{matrix $V$} defines a unitarity triangle (UT) in the complex plane. 
\CP violation measurements and semileptonic decay rates (and other methods)
can be conveniently displayed and compared as constraints on the angles and sides, respectively, of this triangle.
Inconsistencies between all these (in general) precise and redundant constraints can be used to search for new physics (NP). 
As today, there is an impressive overall agreement between all measurements~\cite{ref:globalCKMfits}.
Among these the angle $\gamma$, defined as the phase of $V_{ub}$ in the Wolfenstein parametrization~\cite{ref:CKM},
is particularly relevant since it is the only \CP-violating measurement that,
together with the determination of the \CP-conserving magnitude of $V_{ub}$,
selects a region of the UT apex independently of most types of NP, and thus constitutes a SM candle type of measurement.
Current constraints, provided by the \babar\ and Belle experiments, make use of 
$\Bpm \to D^{(*)} \Kpm$ and $\Bpm \to D \Kstarpm$ decays, and are still weak ($\sim 15^\circ$).
Neutral \B decays have also been proposed, although 
do not yet provide significant constraints.

The angle $\gamma$ from $\Bpm \to D^{(*)} \Kpm$ and $\Bpm \to D \Kstarpm$ decays is determined
measuring the interference between the amplitudes $\b \to \u$ and $\b \to \c$,
when the neutral \D meson is reconstructed in a final state accessible from both \Dz and \Dzb decays. 
Since both amplitudes are tree level, the interference is unaffected by NP appearing in the loops, making the theoretical 
interpretation of observables in terms of $\gamma$ very clean. 
The disadvantage is that the branching fractions of the involved decays are small due to CKM suppression ($10^{-5}-10^{-7}$), 
and the size of the interference, given by the ratio $r_\B$ between the magnitudes of the $\b \to \u$ and $\b \to \c$ amplitudes, 
is small due to further CKM and color suppressions ($\sim 10\%$). As a consequence, the measurements are
statistically limited and one has to combine complementary methods applied on the same \B decay modes sharing the
hadronic parameters ($r_\B$ and $\delta_\B$, i.e. the relative magnitude and phase of the $\b\to\u$ and $\b\to\u$ transitions) and $\gamma$, 
and use as many as possible different \B decay modes to improve the overall sensitivity to $\gamma$.

In this talk we present the most recent determinations of $\gamma$ obtained by \babar, based on the full
data sample of charged \B meson decays produced in $e^+e^- \to \FourS \to \BpBm$ and 
recorded in the years 1999-2007, about $468\times10^6$ \BpBm pairs.
We have studied $\Bpm \to D^{(*)} \Kpm$ and $\Bpm \to D \Kstarpm$ decays,
with the neutral \D mesons reconstructed in a number of different final states: $\D \to \KS h^+ h^-$, with $h=\pi,\K$
(Dalitz plot method); $\D \to \Kpm\pimp$ (ADS method); 
and $\D \to f_{\CP}$, with $f_{\CP}$ a \CP-eigenstate (GLW method)~\cite{ref:dalitz_ads_glw}.



One of the charged \B mesons produced in the \FourS decay is fully reconstructed, with efficiencies ranging 
between 40\% (for low-multiplicity decays with no neutrals) 
and 10\% (for high-multiplicity decays with neutrals). The selection is optimized to maximize the statistical sensitivity. 
The reconstruction efficiencies have substantially improved (20\% to 60\% relative) with respect to our previous measurements 
based on $384\times10^6$ \BpBm pairs, reflecting improvements in tracking and particle identification,
and optimization of analysis procedures.
Signal \B decays are characterized by means of two nearly independent kinematic variables exploiting the constraint from the
known beam energies: 
the beam-energy $\mes \equiv \sqrt{E^{*2}_{\rm beam}-|p^{*}_{\B}|^2}$
and the energy-difference $\DeltaE \equiv E^*_\B - E^*_{\rm beam}$.
Since the main source of background comes from \qqbar continuum production, additional discrimination
is achieved using multivariate analysis tools, from the combination (either a linear Fisher discriminant \fis,
or a non-linear neural network $NN$) of several event-shape quantities. These variables distinguish between
spherical \BB events from more 
jet-like continuum events and exploit the different angular correlations in the two event categories. 
The signal is finally separated from background through 
unbinned maximum likelihood (UML) fits to the
$\Bpm \to D^{(*)} \Kpm$ and $\Bpm \to D \Kstarpm$ data using \mes, \DeltaE, and \fis or $NN$. 
$\Bpm \to D^{(*)} \pipm$ decays, 
which are
about 12 times more abundant than $\Bpm \to D^{(*)} \Kpm$, have a similar topology but are 
discriminated by means of excellent pion and kaon identification provided by $dE/dx$ and Cerenkov measurements,
and show negligible \CP-violating effects ($r_\B \sim 1\%$). Therefere, these decays provide powerful calibration
and control samples for negative tests of \CP violation.


In the Dalitz plot (DP) method the amplitude for a \Bm decay has for the $\b\to\c$ transition the DP of the \Dz decay, 
while for the $\b\to\u$ transition the DP is the corresponding to the \Dzb decay. If we assume no \D mixing nor \CP violation 
in the \D decay, and use as independent kinematic variables $s_\pm=m^2(\KS\pipm)$,
then the two DPs are one rotated $90^\circ$ to each other.
This is of critical importance since allows to determine directly from data the strong charm phase variation 
for \Dz and \Dzb, as well as well as the hadronic parameters $r_\B$ and $\delta_\B$, and the weak phase $\gamma$, provided
that a \D decay amplitude model is assumed. For \Bp decays one has to interchange the \Dz and \Dzb DPs, 
and change the sign of $\gamma$. This results in an interference term proportional to our observables 
$x_\pm \equiv r_\B \cos(\delta_\B \pm \gamma)$ 
and
$y_\pm \equiv r_\B \sin(\delta_\B \pm \gamma)$, i.e.
the real and imaginary parts of the ratio of $\b\to\u$ and $\b\to\c$ amplitudes for \Bpm decays.
%
%
We reconstruct $\Bpm \to D \Kpm$, $D^* \Kpm$ with $D^* \to D\piz,D\gamma$, and $\Bpm \to D \Kstarpm$ with $\Kstarpm \to \KS\pipm$ decays,
followed by neutral \D meson decays to the 3-body self-conjugate final states $\KS h^+ h^-$, with $h=\pi,\K$. 
From the UML fit we determine the signal and background yields in each of the eight different final states for each \B charge, 
along with the \CP-violating 
parameters $x_\pm$ and $y_\pm$~\cite{ref:babar_dalitzpub2010}.
We find 1507 \Bpm signal candidates with $\KS\pip\pim$, and 268 with $\KS\Kp\Km$. 
Prior to the \CP fit, we model the \Dz and \Dzb decay amplitudes as a coherent sum of S-, P-, and D-waves,
and determine their amplitudes and phases (along with other relevant parameters)
relative to the dominant two-body \CP-eigenstates $\KS \rho(770)$ (for $\KS\pip\pim$) and $\KS a_0(980)$ (for $\KS\Kp\Km$),
using a large ($\approx 6.2\times10^5$) and very pure ($\approx 99\%$) signal sample of flavor tagged 
neutral \D mesons from $\Dstarp\to\Dz\pip$ decays produced in $e^+e^- \to\c\cbar$ events~\cite{ref:dmixing-kshh}.
%
%
\begin{figure}[htb!]
\begin{center}
\vskip-0.1cm
\begin{tabular}{ccc}
\includegraphics[width=0.32\textwidth]{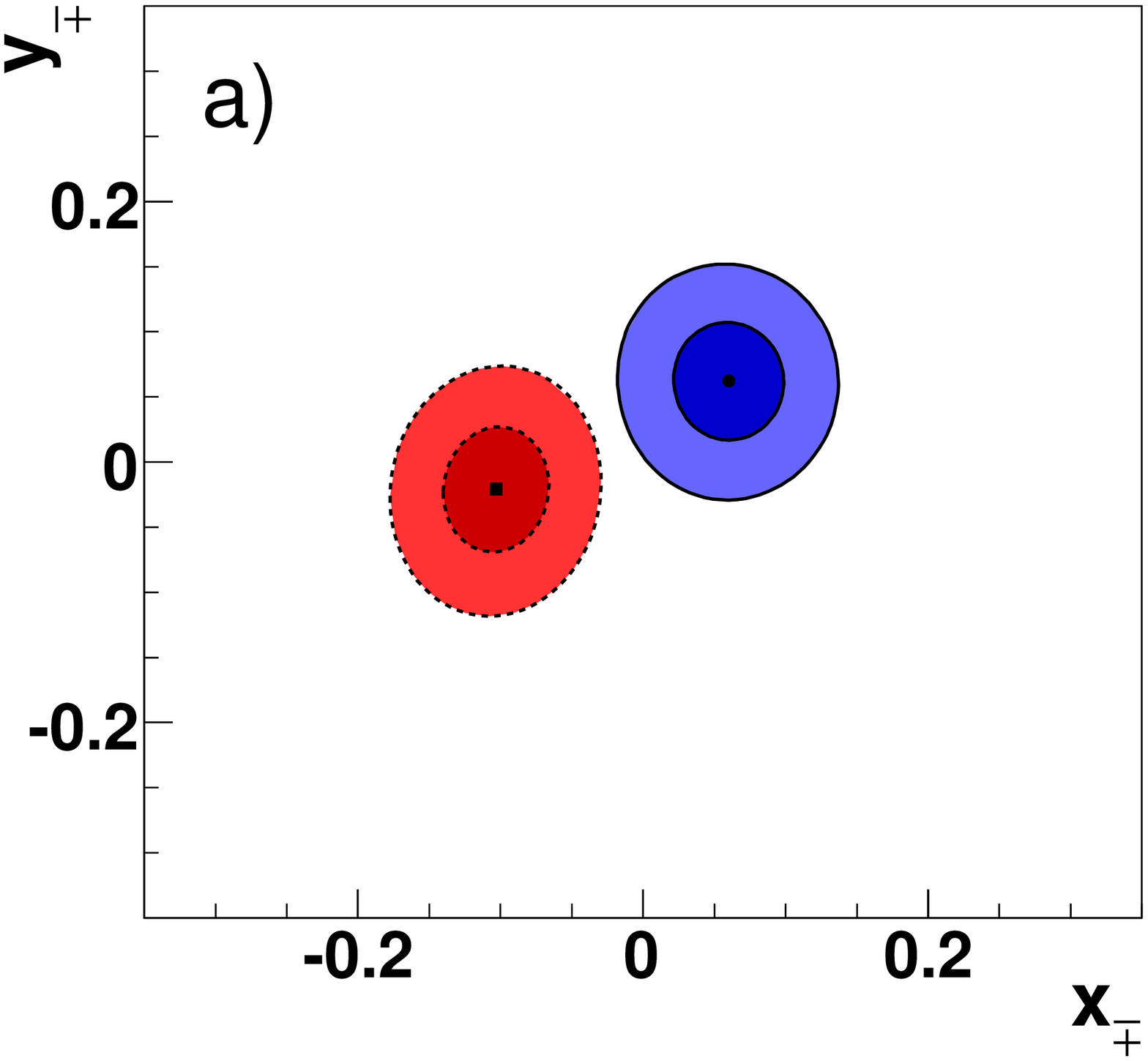}&
\includegraphics[width=0.32\textwidth]{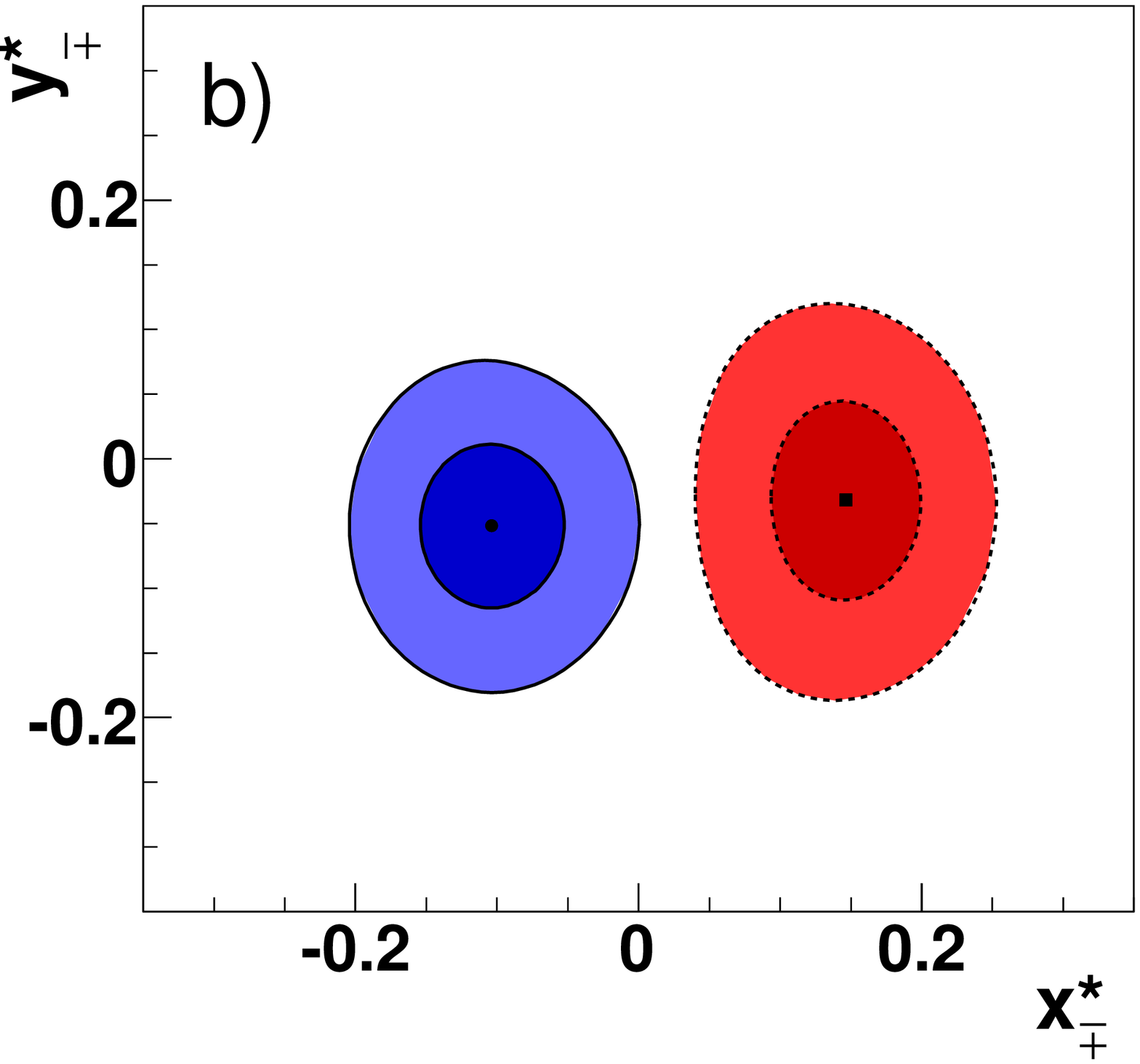} &
\includegraphics[width=0.32\textwidth]{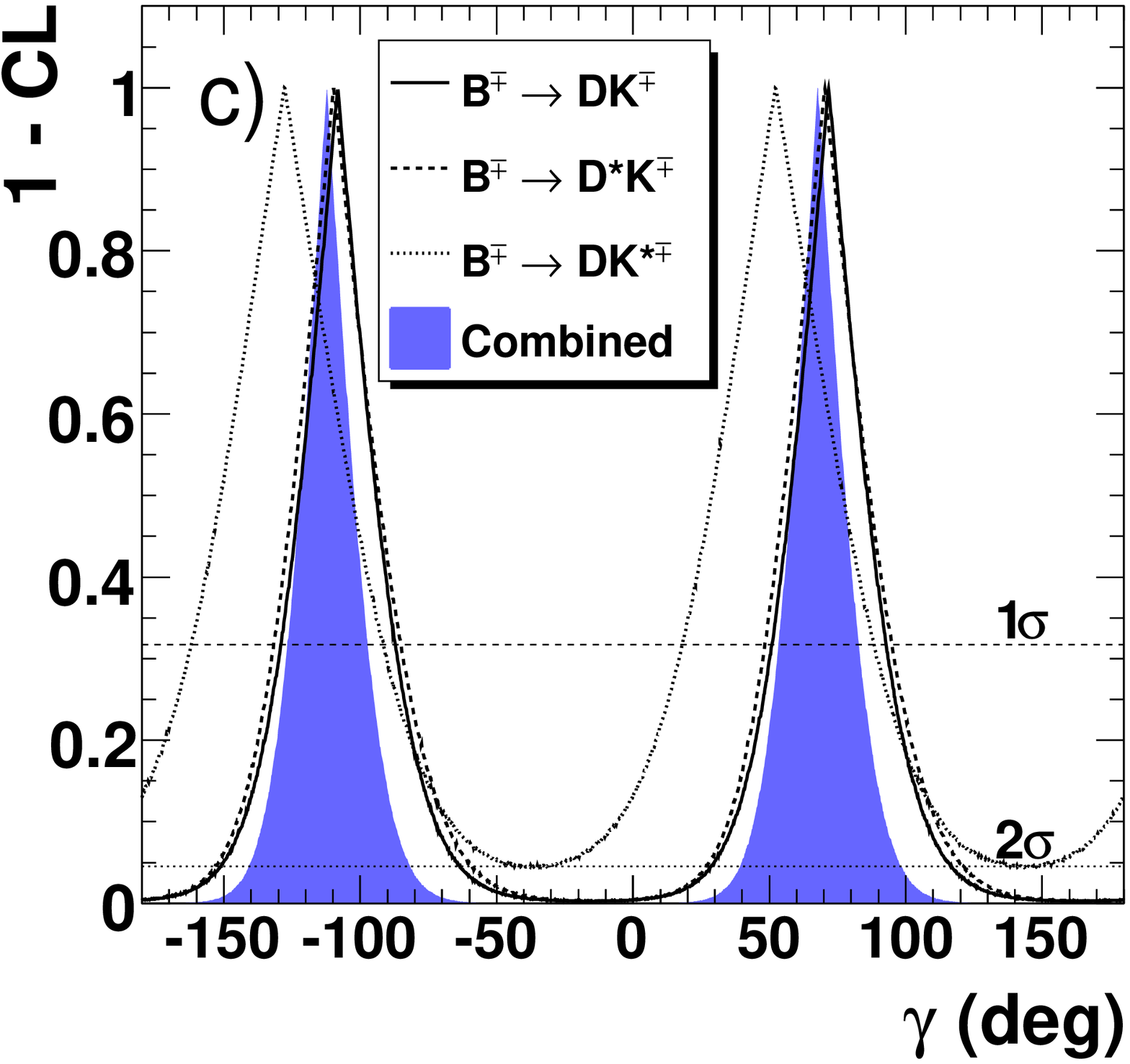} \\
\end{tabular}
\end{center}
\caption{\label{fig:dalitz} $1\sigma$ and $2\sigma$ contours in the $(x_\pm,y_\pm)$ planes for
(a) $\Bpm \to D \Kpm$ and (b) $\Bpm \to D^* \Kpm$, for \Bm (solid lines) and \Bp (dotted lines) decays.
(c) $1 - {\rm CL}$ as a function of $\gamma$ for $\Bpm \to \D\Kpm, \Dstar\Kpm, \D\Kstarpm$ decays.
The dashed (upper) and dotted (lower) horizontal lines correspond to the $1\sigma$ and $2\sigma$ intervals, respectively.
}
\end{figure}
From the $(x_\pm,y_\pm)$ confidence regions for each of the 3 different \B decay modes --Fig.~\ref{fig:dalitz}.(a)(b)-- we determine, 
using a frequentist procedure, $1\sigma$ $[2\sigma]$ intervals for $\gamma$ --Fig.~\ref{fig:dalitz}.(c)--. 
We obtain $\gamma~({\rm mod}~180^\circ) = (68\pm14\pm4\pm3)^\circ$ $[39^\circ,98^\circ]$, where the three uncertainties are statistical,
experimental systematic, and amplitude model systematic.
We also determine the hadronic parameters
$r_\B^{DK^\pm}=(9.6\pm2.9)\%$ $[3.7,15.5]\%$, 
$r_\B^{D^*K^\pm}=(13.3^{+4.2}_{-3.9})\%$ $[4.9,21.5]\%$, 
$\kappa r_\B^{DK^{*\pm}}=(14.9^{+6.6}_{-6.2})\%$ $[0,28.0]\%$ ($\kappa = 0.9\pm0.1$ takes into account 
the \Kstar intrinsic width),
and the strong phases 
$\delta_B^{DK^\pm}$, 
$\delta_B^{D^*K^\pm}$, 
and $\delta_B^{DK^{*\pm}}$~\cite{ref:babar_dalitzpub2010}. 
A $3.5\sigma$ evidence of direct \CP violation ($\gamma \ne 0$) is found from the combination of the 3 channels,
which corresponds to the significance of the separation between the $(x_+,y_+)$ and $(x_-,y_-)$ solutions in Fig.~\ref{fig:dalitz}.(a)(b).

\begin{figure}[htb]
\begin{center}
\vskip-0.1cm
\begin{tabular}{cc}
\includegraphics[width=0.45\textwidth]{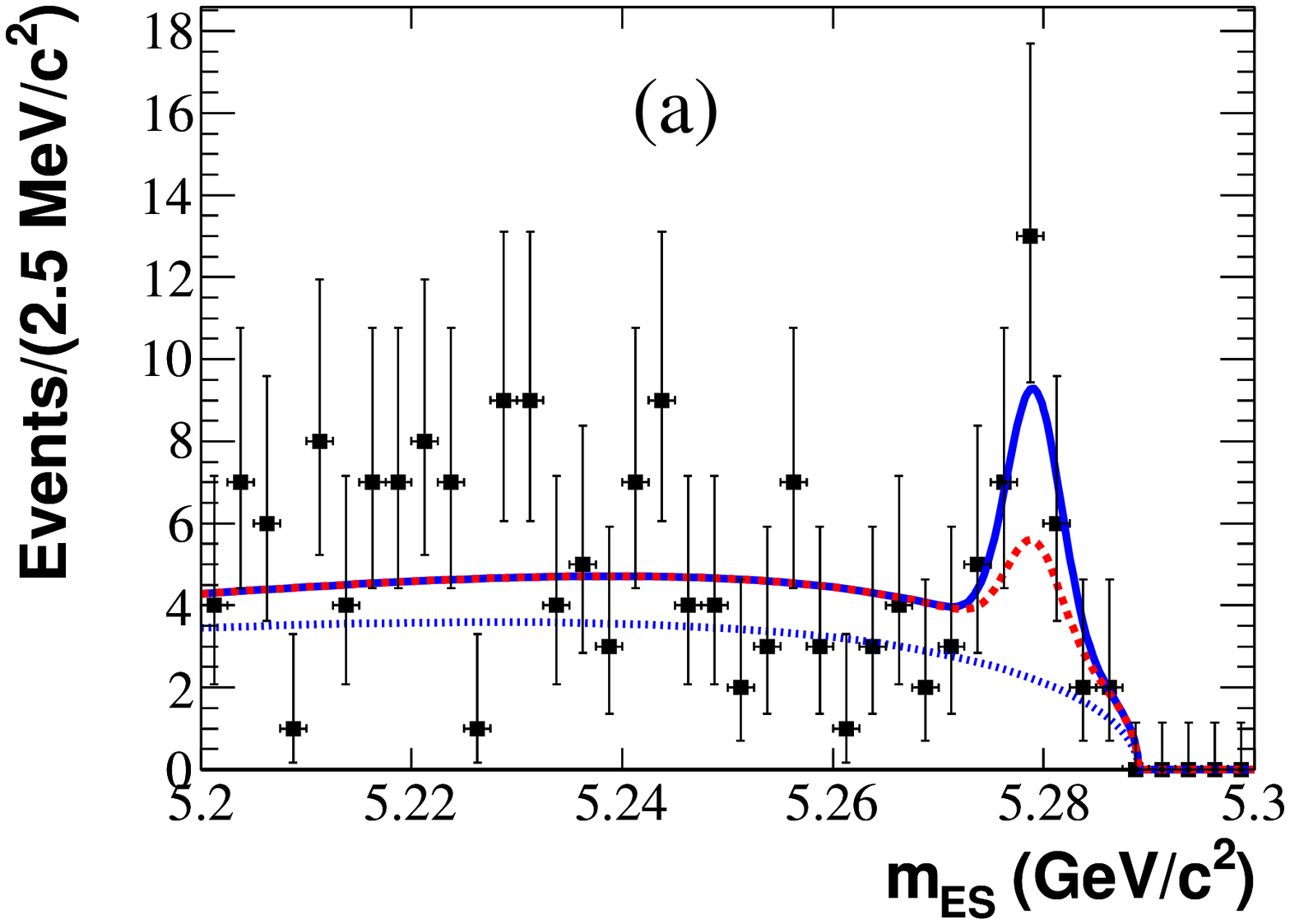} &
\includegraphics[width=0.45\textwidth]{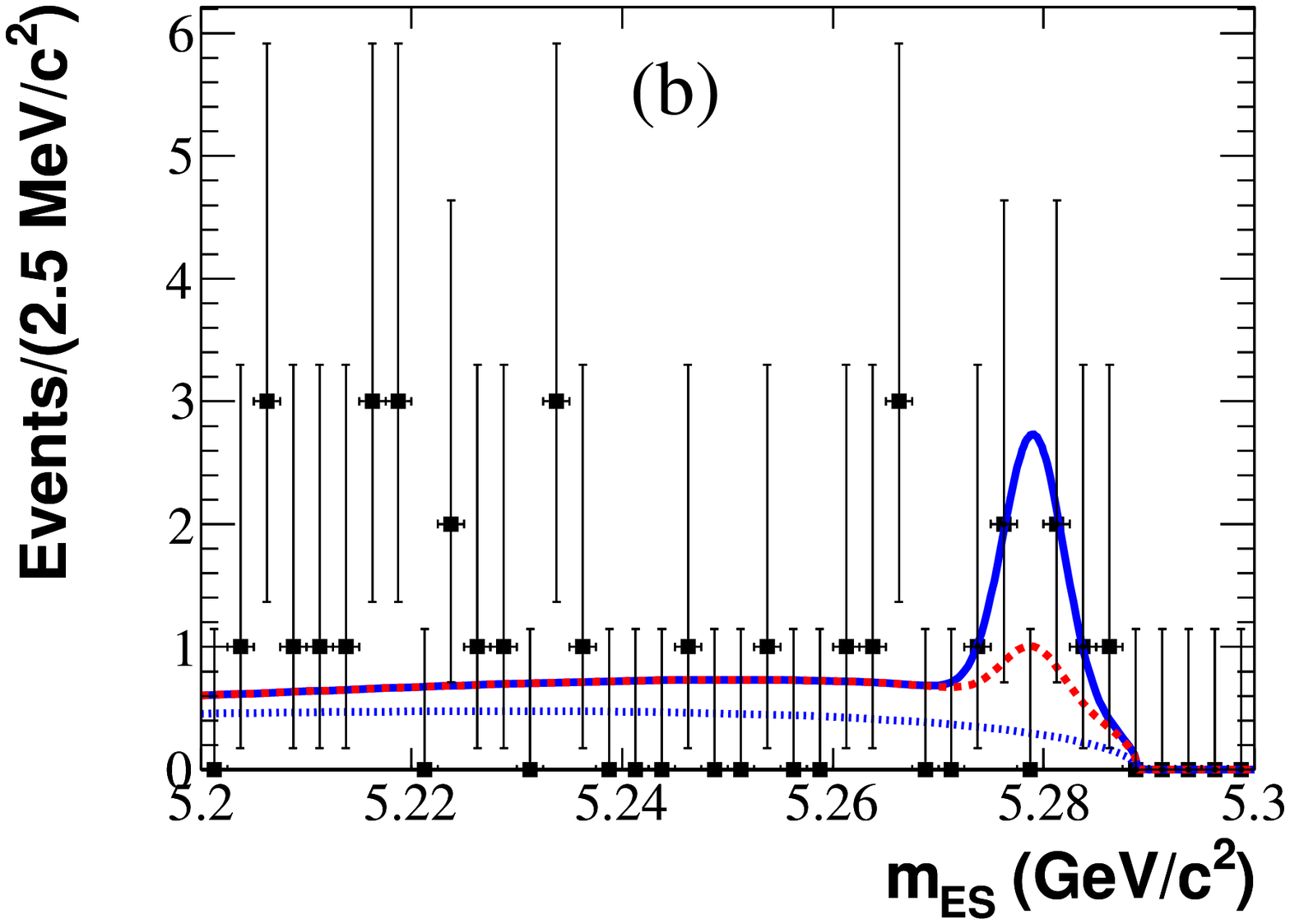} \\
\end{tabular}
\end{center}
\caption{\label{fig:ads}
Projections on \mes for (a) $\Bpm \to \D\Kpm$ and (b) $\Bpm \to D^*[\D\piz]\Kpm$, $\D\to\Kmp\pipm$ 
opposite-sign decays, for ADS samples enriched in signal ($NN>0.94$). 
The points with error bars are data while the curves represent the fit projections for
signal plus background (solid), the sum of all background components (dashed), and $q\bar q$ background only (dotted).
}
\end{figure}

In the ADS method, we reconstruct $\Bpm \to D \Kpm$, $D^* \Kpm$ with $D^* \to D\piz,D\gamma$, followed by \D decays to both the doubly-Cabibbo-suppressed (DCS)
\Dz final state $K^+\pi^-$ and the Cabibbo-favored (CF) $K^-\pi^+$, which is used as normalization and control sample. 
Final states with opposite-sign kaons arise 
either from the CKM favored \B decay followed by the DCS \D decay
or from the CKM- and color-suppressed \B decay followed by the CF \D decay,
producing an interference which can be potentially large since the magnitudes of the interfering amplitudes are similar.
However, their overall branching ratios are very small ($\sim 10^{-7}$) and background suppression becomes crucial.
The UML fit directly determines the three branching fraction ratios $R_{ADS}$ between \B decays with opposite-sign 
and same-sign kaons, and the three yields of \B decays with same-sign kaons, using \mes and $NN$.
The three \CP asymmetries $A_{ADS}$ are inferred from all these. 
We obtain first indications of signals for the
$\Bpm \to D \Kpm$ and $\Bpm \to D^* \Kpm$ (with $D^* \to D\piz$) opposite-sign modes --Fig.~\ref{fig:ads}--, 
with significances of $2.1\sigma$ and $2.2\sigma$, respectively~\cite{ref:babar_ADS-DK-DstarK}.
The measured branching fraction ratios are 
$R_{ADS}^{DK}=(1.1\pm0.5\pm0.2)\times10^{-2}$, 
$R_{ADS}^{[D\piz]K}=(1.8\pm0.9\pm0.4)\times10^{-2}$, and
$R_{ADS}^{[D\gamma]K}=(1.3\pm1.4\pm0.8)\times10^{-2}$, and the \CP asymmetries are
$A_{ADS}^{DK}=-0.86\pm0.47^{+0.12}_{-0.16}$, 
$A_{ADS}^{[D\piz]K}=0.77\pm0.35\pm0.12$, and
$A_{ADS}^{[D\gamma]K}=0.36\pm0.94^{+0.25}_{-0.41}$.
From these results and external measurements of the relative amplitude and phase of \Dzb to \Dz mesons 
decaying into the $K^- \pip$ final state~\cite{ref:hfag} we infer, using a frequentist procedure similar to that used in
the DP method,
$r_\B^{DK^\pm}=(9.5^{+5.1}_{-4.1})\%$ $[0,16.7]\%$, $r_\B^{D^*K^\pm}=(9.6^{+3.5}_{-5.1})\%$ $[0,15.0]\%$,
and the strong phases $\delta_B^{DK^\pm}$, $\delta_B^{D^*K^\pm}$, in good agreement with those obtained with the DP technique.
%
%



In the GLW method, we reconstruct $\Bpm \to D \Kpm$ decays, followed by \D decays to non-\CP ($\Dz\to\Km\pip$), \CP-even ($\Kp\Km$, $\pip\pim$), 
and \CP-odd ($\KS\piz$, $\KS\phi$, $\KS\omega$) eigenstates. The partial decay rate charge asymmetries $A_{\CP\pm}$ for
\CP-even and \CP-odd \D final states and the ratios $R_{\CP\pm}$ of the charged-averaged \B meson partial decay
rates in \CP ($R_{K/\pi}^\pm$) and non-\CP ($R_{K/\pi}$) decays (normalized to the corresponding $\Bpm \to D \pipm$ decays, 
to reduce systematic uncertainties)
provide four observables from which the three 
unknowns $\gamma$, $r_\B$ and $\delta_\B$ can be extracted (up to an 8-fold ambiguity for the phases). 
The signal yields, expressed in terms of $A_{\CP\pm}$, $R_{K/\pi}^\pm$ and $R_{K/\pi}$ are extracted from UML fits 
to \mes, \DeltaE, and \fis. We identify about 500 $\Bpm \to D \Kpm$ decays with \CP-even \D final states and a similar
amount for \CP-odd \D final states, and measure~\cite{ref:babar_GLW-DK} 
$A_{\CP+}=0.25\pm0.06\pm0.02$, 
$A_{\CP-}=-0.09\pm0.07\pm0.02$,
$R_{\CP+}=1.18\pm0.09\pm0.05$, and
$R_{\CP-}=1.07\pm0.08\pm0.04$. 
The parameter $A_{\CP+}$ is different from zero with a significance of $3.6\sigma$, and constitutes evidence for
direct \CP violation in $\Bpm \to D \Kpm$ decays. 
These results can be written in terms of the observables
$x_\pm$ using the relationship $x_\pm = [R_{\CP+}(1\mp A_{\CP+}) - R_{\CP-}(1\mp A_{\CP-})]/4$.
Excluding the $D\to\KS\phi$, $\phi\to K^+K^-$ channel to facilitate the combination with the DP method,
we find $x_+=-0.057\pm0.039\pm0.015$ and $x_-=0.132\pm0.042\pm0.018$, which are consistent (and of similar
precision) with the DP method.
From these results and using a frequentist procedure similar to that used previously we infer
$24\% < r_\B < 45\%$ $[6,51]\%$, and
mod $180^\circ$, 
$11^\circ < \gamma < 23^\circ$ or $81^\circ < \gamma < 99^\circ$ or $157^\circ < \gamma < 169^\circ$
$[7^\circ,173^\circ]$.


We have reported the recent progress in the determination of the CKM angle $\gamma$, using the complete \babar\ data sample and three different
and complementary methods (DP, ADS, and GLW). 
A coherent and consistent set of results on $\gamma$ and the hadronic parameters characterizing the \B decays has been
obtained. The central value for $\gamma$, around $70^\circ$ with a precision around $15^\circ$, is consistent with indirect determinations from
CKM fits~\cite{ref:globalCKMfits}. 
A proper average of all the three methods using the full \babar\ sample of $\Bpm\to D^{(*)} K^{\pm}$, $D K^{*\pm}$ decays is foreseen.
We obtain $x_- - x_+ = 0.175\pm0.040$ by combining the $x_\pm$ measurements from the DP and GLW methods for $\Bpm\to D K^{\pm}$ decays,
which is different from zero with a significance of $4.4\sigma$,
thus constitutes strong evidence for direct \CP violation in these charged \B decays.
Finally, we have the first sign of an ADS signal in $\Bpm\to D K^{\pm}$ and  $\Bpm\to D^{(*)} K^{\pm}$ decays.

\end{document}

%% file: mysymbols.tex
\newcommand{\fis}{\ensuremath{\mbox{$\mathcal{F}$}}\xspace}

\def\beq{\begin{equation}}
\def\eeq{\end{equation}}
\def\bea{\begin{eqnarray}}
\def\eea{\end{eqnarray}}
\def\bq{\begin{quote}}
\def\eq{\end{quote}}
\def\ben{\begin{enumerate}}
\def\een{\end{enumerate}}

\def\D       {\ensuremath{D}\xspace}
\def\K  {\ensuremath{K}\xspace}
\newcommand{\kevc}{\ensuremath{{\mathrm{\,ke\kern -0.1em V\!/}c}}\xspace}
\newcommand{\kevcc}{\ensuremath{{\mathrm{\,ke\kern -0.1em V\!/}c^2}}\xspace}
\def\figurebox#1#2#3{%
    \def\arg{#3}%
    \ifx\arg\empty
    {\hfill\vbox{\hsize#2\hrule\hbox to #2{\vrule\hfill\vbox to #1{\hsize#2\vfill}\vrule}\hrule}\hfill}%
    \else
    {\hfill\epsfbox{#3}\hfill}%
    \fi}